\documentclass[10pt]{iopart}
\usepackage{epsf}
\usepackage{graphicx}
\begin{document}

\title{Symmetry breaking in the self-consistent Kohn-Sham equations}


\author{E. Prodan}
\address{University of California, Department of Physics, Santa
Barbara, CA 93106, USA.}

\begin{abstract}
  The Kohn-Sham (KS) equations determine, in a self-consistent way, the
  particle density of an interacting fermion system at thermal equilibrium.
  We consider a situation when the KS equations are known to have a
  unique solution at high temperatures and that this solution is a
  uniform particle density. We prove that, at zero temperature, there
  are stable solutions that are not uniform. We provide the general
  principles behind this phenomenon, namely the conditions when it
  can be observed and how to construct these non-uniform solutions.
  Two concrete examples are provided, including fermions on the
  sphere which are shown to crystallize in a structure that
  resembles the C$_{60}$ molecule.
\end{abstract}

\section{Introduction}

We consider a system of $N$ interacting fermions in a finite volume.
Since we want to avoid the surface effects, we actually consider the
fermions moving on toruses and spheres, or, more generally, on a
closed Riemann manifold ${\cal M}$ of finite volume $\Omega$.
According to the Kohn-Sham theory \cite{Kohn1} (and later extensions
\cite{Mermin}), the particle density at thermal equilibrium at a
temperature $T$ ($\beta=1/kT$) is a solution of the following set of
equations:
\begin{equation}\label{KS1}
     (H_0+V_n)\phi_i = \epsilon_i \phi_i
\end{equation}
\begin{equation}\label{KS2}
     n(x) = \sum_i \left
     (1+e^{\beta(\epsilon_i-\mu)}\right)^{-1}|\phi_i(x)|^2,
\end{equation}
with $V_n$ an effective potential depending entirely on the particle
density $n$ and $\mu$ determined from $\int n(x)dx=N$. $H_0$ is the
single particle, non-interacting Hamiltonian. We refer to
\begin{equation}
H_n\equiv H_0+\lambda V_n
\end{equation}
as the Kohn-Sham Hamiltonian, where we introduced the coupling
constant $\lambda$ for convenience. We will neglect the spin degree
of freedom.

There is no closed form of $V_n$. However, at least for the electron
gas, there is a set of very successful, explicit approximations,
which already provide numerical results that are within the so
called "chemical precision" \cite{Kohn2}. Although in this paper we
do not use a specific approximation, we will often make reference to
two such approximations in order to check our assumptions. They are
the Local Density Approximation (LDA): $V_n=v*n+v_{xc}(n)$ ($v$ the
two body interaction), i.e. $V_n$ is the sum between the Hartree
potential and a {\it function} of density, and the quadratic
approximation (QA): $V_n=K*n$, i.e. $V_n$ is the convolution of the
density with a certain kernel. The mathematical structure of QA is
the same as that of the Hartree approximation.

In this paper, we are not concerned with the physical and
mathematical principles leading to the KS equations, but rather with
the mathematical structure of these equations, in particular with
the question of uniqueness at zero temperature. In other words, we
have already assumed $v$-representability, picked our approximation
for $V_n$ and we are ready to compute the self-consistent solutions.
What should we expect? Well, as already demonstrated (see for
example Ref.~\cite{Reimann}), even for local approximations, we
should expect a very rich structure, which may include multiple
solutions, symmetry breaking etc.. While previous studies used
purely numerical methods, here we use group theoretical methods and
functional analysis to study this structure.

Let us discuss first what it is known about the KS equations at
finite temperature and finite volume. The notation $\|\phi\|_{L^p}$
stands for $[\int_{\cal M}|\phi(x)|^pdx]^{1/p}$ and $\phi \in L^p$
means $\|\phi\|_{L^p}<\infty$. Assume the following:
\begin{description}
    \item[A1)] $H_0$ is self adjoint, bounded from below; for $-a$ below its energy spectrum,
    the kernel $(H_0+a)^{-2}(x,x')$ is continuous (with respect to $x$ and $x^\prime$) and
    \begin{equation}
    k_a\equiv\sup_{x\in{\cal M}}(H_0+a)^{-2}(x,x)<\infty.
    \end{equation}
    \item[A2)] $V_n\in L^2({\cal M})$ and
    $w\equiv \sup \|V_n\|_{L^2}<\infty$, where the supremum is taken
    over all $n$ in
    \begin{equation}
    S^N\equiv\{n\in L^1({\cal M}),\ \|n\|_{L^1}=N\}.
    \end{equation}
\end{description}
As $H_0$ is in general equal to minus the Laplace operator, A1 is
easy to check for 1, 2 and 3 dimensional toruses or spheres. It
fails in 4 and higher dimensions. Since ${\cal M}$ is of finite
volume, A1 automatically implies that $\exp(-\beta H_0)$ is trace
class. A1 also implies that $\|f\|_{L^{\infty}}\leq
\sqrt{k_a}\|(H_0+a)f\|_{L^2}$. Together with A2 (easy to verify for
LDA and QA \cite{Prodan1,Prodan2}), this leads to
\begin{equation}\label{gamma}
    \|V_n(H_0+a)^{-1}\|\leq w\sqrt{k_a}\equiv\gamma_a.
\end{equation}
Then, $H_n$ is self-adjoint for all $n\in S^N$ and, as it follows
from Ref. \cite{Prodan1}, the Kohn-Sham equations can be formulated
as a fixed point problem:

\medskip

\noindent {\bf Theorem 1.} For $T>0$, the following map is well
defined:
\begin{eqnarray}\label{Tmap}
    \textsf{T}: S^N\rightarrow S^N \nonumber \\
    S^N\ni n\rightarrow
    \textsf{T}[n](x)=\left(1+e^{\beta(H_n-\mu)}\right)^{-1}(x,x),
\end{eqnarray}
where $\mu$ is the unique solution of
$N=Tr\left(1+e^{\beta(H_n-\mu)}\right)^{-1}$. The fixed points of
$\textsf{T}$ generate all possible solutions of the KS equations.

\medskip

Many will recognize in Eq.~(\ref{Tmap}) the usual formulation of the
KS problem in terms of the density matrix. When appealing to the
fixed point theorem, the functional form of the map $\textsf{T}$ and
its domain of definition are {\it equally} important. What is new in
the above result is that $\textsf{T}$ is well defined for all
densities which integrate to $N$.

Apart from complications that may occur at low particle densities
and which will not be addressed here, the following assumption can
be easily verified for LDA and QA (see Ref. \cite{Prodan1,Prodan2}):
\begin{description}
    \item[A3)] There exists $\chi<\infty$ such that
    \begin{equation}
    \|V_n-V_{n'}\|_{L^1}\leq \chi\|n-n'\|_{L^1},
    \end{equation}
    for any $n$, $n'\in S^N$.
\end{description}
If A1-A3 are satisfied, then there exists $\kappa$, which is a
function of $\lambda$, such that \cite{Prodan1}
\begin{equation}
    \|\textsf{T}[n]-\textsf{T}[n^\prime]\|_{L^1}\leq
    \kappa\|n-n'\|_{L^1},
\end{equation}
on the entire $S^N$. For $\lambda$ smaller than a critical value
$\lambda_c$, $\textsf{T}$ becomes a contraction and, consequently,
it has a unique fixed point. If the constants can be chosen
independent of temperature in A1-A3, it is not hard to show that
$\lambda_c$ increases with temperature. In other words, if $\lambda$
is kept fixed, A1-A3 (and the fact that $\Omega$ is finite)
guaranties the existence of a unique fixed point of $\textsf{T}$ at
high temperatures.

Let us end the finite temperature case with a few remarks. For an
exact $V_n$, the existence and uniqueness, at any finite $T$, will
follow from the convexity of the functional \cite{Levy}, provided
that the equilibrium density can be written as in Eq.~(\ref{KS2})
(i.e. is $v$-representable). In practice, we don't have the exact
$V_n$ and $v$-representability has not been yet proved or disproved.
Also, in the thermodynamic limit, where systems can have multiple
coexisting phases, the issue of uniqueness becomes more delicate and
definitely there are many opened questions here. Thus, the question
of existence and uniqueness in the finite temperature Kohn-Sham
equations is not trivial.

The situation at zero temperature is more delicate. The density now
becomes $n(x)=\sum |\phi_i(x)|^2$ where the sum goes over the lowest
$N$ energy states of $H_n$. If the last occupied energy level is
degenerate and only partially occupied, there is an ambiguity in
defining $n(x)$. In this paper, we deal exactly with this situation.

Let us assume that that there is a continuous group $G_c$ acting
ergodically on ${\cal M}$ and preserving the Riemann structure. On
torus or sphere, this group will be simply the translations or
rotations. Let us consider the natural unitary representation of
$G_c$ in $L^2({\cal M})$:
\begin{equation}
    G_c\ni g\rightarrow \hat{g},\ (\hat{g}f)(x)=f(gx).
\end{equation}
We assume that $H_0$ commutes with all $\hat{g}$ and that every
symmetry of the particle density is automatically a symmetry of the
effective potential:
\begin{description}
    \item[A4)] If $n(gx)=n(x)$, then $V_n(gx)=V_n(x)$
    (equivalently $\hat{g}V_n\hat{g}^{-1}=V_n$).
\end{description}
This assumption can be easily verified for LDA and QA. Besides other
things, A4 implies that $V_n$ is a constant if $n(x)$ is uniform,
and we can fix this constant to zero. In other words, the Kohn-Sham
Hamiltonian reduces to $H_0$ if $n(x)=\bar{n}$ ($\bar{n}=N/\Omega$).
Then, it is trivial to show that, at any finite temperature,
$\bar{n}$ is a solution of the KS equations. With our assumptions,
we also know that this is the only solution at high temperatures. At
zero temperature, assume that, if we populate with $N$ particles the
energy levels of $H_0$, from smaller to higher energies, we end up
with $N_0$ particles on the last occupied energy level, assumed
$d$-fold degenerate with $d>N_0$. We refer to this level and its
energy as the Fermi level and Fermi energy $\epsilon_F^0$. If we can
find $N_0$ states at the Fermi level so as to generate a uniform
particle density, then $\bar{n}$ is a solution of the KS equations.
If there is no such combination of states, than either there is no
solution, the solution is not uniform or we need to consider
fractional occupation numbers. We will not discuss here the last
possibility, but rather concentrate to the $T\rightarrow 0$ limit of
Eqs.~(\ref{KS1}) and (\ref{KS2}).

We now show when and how the non-uniform solutions can be found. We
look for a finite subgroup $G$ of $G_c$, which has to satisfy two
simple conditions. We index its irreducible representations by
$\Gamma$ and use the symbol $|\Gamma|$ to specify their dimension.
Let $P_\Gamma$ denote the projectors
\begin{equation}
    P_\Gamma=\frac{|\Gamma|}{|G|}\sum \limits_{g\in G}
    \chi_\Gamma(g)\hat{g},
\end{equation}
with the following properties
\begin{equation}
P_\Gamma P_{\Gamma\prime}=\delta_{\Gamma\Gamma\prime}, \ \
\sum_\Gamma P_\Gamma=I.
\end{equation}
Above, $|G|$ denotes the cardinal of $G$ and $\chi_\Gamma (g)$ the
character of $g$ in the representation $\Gamma$. Let ${\cal H}_F$
denote the eigenspace of $H_0$ corresponding to $\epsilon_F^0$. This
space is invariant to $G_c$ and it decomposes according to the
irreducible representations of $G$: ${\cal H}_F=\oplus_i
P_{\Gamma_i}{\cal H}_F$. The sum goes only over those $\Gamma$ for
which $P_{\Gamma}{\cal H}_F \neq 0$. In general, $\dim
P_{\Gamma_i}{\cal H}_F =n_i |\Gamma_i|$, where $n_i$ is the number
of representations $\Gamma_i$ on ${\cal H}_F$. The subgroup $G$ we
look for must satisfy the following:
\begin{description}
    \item[A5)] $\dim P_{\Gamma_i}{\cal H}_F=|\Gamma_i|$, i.e. we
    have irreducible representations of $G$ in each $P_{\Gamma_i}{\cal
    H}_F$.
    \item[A6)] $|\Gamma_i|=N_0$ for some $i$ (we rearrange
    so that $i=0$).
\end{description}
Let now $n_0(x)$ be the particle density when we populate all the
states of $H_0$ below $\epsilon_F^0$ plus the $N_0$ states in
$P_{\Gamma_0}{\cal H}_F$. Since we assumed that $\dim {\cal
H}_F>N_0$, $n_0(x)$ is not uniform. The last condition is on the
effective potential:
\begin{description}
    \item[A7)] If $V_{\Gamma_i}\equiv \langle
    \phi^0_{\Gamma_i},V_{n_0}\phi^0_{\Gamma_i}\rangle$ with
    $\phi^0_{\Gamma_i}$ any norm one vector from $P_{\Gamma_i}{\cal H}_F$, then
    \begin{equation}
    V_{\Gamma_i}-V_{\Gamma_0}>0,  \textrm{for all} \ i>0.
    \end{equation}
\end{description}

If the subgroup $G$ satisfying A5-A6 exists and the effective
potential satisfies A7, then, at least for small $\lambda$, the zero
temperature KS equations have a non-uniform solution,
\begin{equation}
    n(x)=n_0(x)+o(\lambda).
\end{equation}
This is our main result.

We end our long introduction with a discussion of the conditions
A5-A7. The assumption A5 greatly simplifies our proof, but it is not
essential (though we don't have a proof without A5). A6 assures the
closed shell condition and is essential. It can be relaxed, for
example, we can have two completely filled shells. However, we
believe that, in the ground state configuration, all $N_0$ particles
occupied the same (lowest) energy level.

Condition A7 refers to the effective potential and it requires,
quite naturally, that the level populated by the $N_0$ particles to
have the lowest energy (in the first order in $\lambda$). We believe
A7 is essential. Now, even if we find the correct $G$, A7 may not be
satisfied, since there is a competition between the Hartree and
exchange-correlation potentials. For repulsive interactions, the
exchange-correlation potential must dominate the Hartree potential
(see Eq.~(\ref{A7})). This is why the electrons crystallize at low
densities. For attractive interactions (like the Lenard-Jones
fermions) is viceversa, the Hartree term has to dominate the
exchange-correlation.

To conclude, A5-A6 determines the crystal structure and A7
determines the conditions, like the range of densities, in which
this structure is stable.

\section{The Proof}

The idea behind our proof is the following. We restrict the search
for $n(x)$ to the densities that are symmetric relative to $G$, and
in a small vicinity of $n_0$. Under the action of $V_n$, the Fermi
level splits into sub-levels, and for $\lambda$ small enough, we
show that, for all densities in this vicinity, there are exactly $N$
states below $\epsilon_F^0$. This allows us to define, for any such
$n$, the density $n^\prime$ corresponding to the potential $V_n$,
i.e. a map $\textsf{T}:\ n\rightarrow n^\prime$. The
self-consistency means $n^\prime=n$, i.e. $n$ is a fixed point for
$\textsf{T}$. To show that $T$ has a fixed point, we prove that this
small vicinity around $n_0$ is mapped into itself by $\textsf{T}$
and that $\textsf{T}$ is a contraction.

To define the space $L^1_{sym}$ of $G$-symmetric densities
precisely, we consider the isometries
\begin{equation}
    T_g: L^1({\cal M})\rightarrow L^1({\cal M}),\ \ (T_gn)(x)=n(gx),
\end{equation}
and define
\begin{equation}
    L^1_{sym}\equiv\bigcap\limits_{g\in G} Ker(T_g-I).
\end{equation}
It is important to notice that $L^1_{sym}$ is a closed subspace of
$L^1({\cal M})$. Since the solutions of the KS equations are not
affected if we add a constant to $V_n$, we can assume without losing
generality that $V_{\Gamma_0}<0$, $V_{\Gamma_i}>0$ for $i>0$ and
$\min\limits_{i>0} V_{\Gamma_i}=|V_{\Gamma_0}|$.

\medskip

\noindent {\bf Theorem 2.} Let us consider the closed subset of
$L^1({\cal M})$,
\begin{equation}
S^{N,\epsilon}_{sym}\equiv \{n\in L^1_{sym},\ \|n\|_{L^1}=N, \
\|n-n_0\|_{L^1}\leq \epsilon\}.
\end{equation}
Then, for $\epsilon$ and $\lambda$ small enough:
\begin{description}
    \item[i)] The following map is well defined
    \begin{equation}
\textsf{T}_\epsilon:S^{N,\epsilon}_{sym}\rightarrow
S^{N,\epsilon}_{sym}, \ \textsf{T}_\epsilon[n](x)=P^<_n(x,x),
\end{equation}
where $P^<_n$ denotes the spectral projector of $H_n$ onto the
spectrum below $\epsilon_F^0$ (excluding $\epsilon_F^0$).
    \item[ii)] $\textsf{T}_\epsilon$ has one and only one fixed point.
    \item[iii)] This fixed point is a solution of the KS equations.
\end{description}

\medskip

\begin{figure}
\begin{center}
\includegraphics[width=10.0cm]{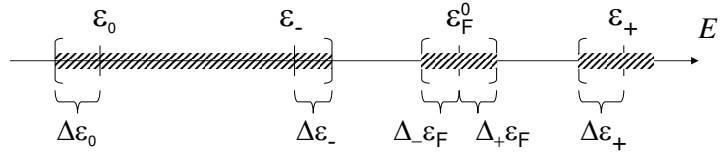}\\
\end{center}
  \caption{If $\epsilon_0$ and $\epsilon_\mp$ denote the ground state and
  energy levels below/above $\epsilon_F^0$  of $H_0$, then, at small
  $\lambda$, the spectrum of $H_n$ ($n\in S^N$) is contained in the hachured
  region of the real axis, where: $\Delta \epsilon_0=\lambda
  \gamma_a (\epsilon_0+a)$, $\Delta \epsilon_-=\frac{\lambda
  \gamma_a}{1-2\lambda \gamma_a} (\epsilon_-+a)$, $\Delta_- \epsilon_F=\lambda
  \gamma_a (\epsilon_F^0+a)$, $\Delta_+ \epsilon_F=\frac{\lambda
  \gamma_a}{1-2\lambda \gamma_a} (\epsilon_F^0+a)$ and $\Delta \epsilon_+=\lambda
  \gamma_a (\epsilon_++a)$. We define ${\cal I}$ as the
hachured region corresponding to some small, fixed $\bar{\lambda}$
and always consider $\lambda<\bar{\lambda}$.
  }
\end{figure}

{\it Proof.} {\it i)} Let us show first that $\textsf{T}_\epsilon$
takes $S^{N,\epsilon}_{sym}$ into $L^1_{sym}$. Since we exclude
$\epsilon_F^0$, $P^<_n$ is well defined for all $n\in
S^{N,\epsilon}_{sym}$ and can be determined from the resolvent of
$H_n$. Also, A1-A2 guaranties that the kernel of $P^<_n$ is
continuous, thus its diagonal is well defined. From A4, $H_n$ and
consequently $P^<_n$ commutes with all $\hat{g}$, $g\in G$, for all
$n\in S^{N,\epsilon}_{sym}$. Then
\begin{equation}
    \textsf{T}_\epsilon[n](gx)=P^<_n(gx,gx)=
    (\hat{g}P^<_n\hat{g}^{-1})(x,x)=\textsf{T}_\epsilon[n](x).
\end{equation}

Next, we show that $\textsf{T}_\epsilon$ takes
$S^{N,\epsilon}_{sym}$ into $S^N$. For this, we need to show that
$H_n$ has exactly $N$ states below $\epsilon_F^0$, for all $n\in
S^{N,\epsilon}_{sym}$. For $\lambda$ small, a first, rough location
of the spectrum can be obtained from Eq. (\ref{gamma}). An
elementary argument will show that the spectrum of $H_n$ is always
located inside the set ${\cal I}$ defined and shown in Fig.~1. We
now investigate the splitting of the Fermi level. For any $n\in
S^{N,\epsilon}_{sym}$, the Fermi level will split in sub-levels,
each corresponding to the different irreducible representations
$\Gamma_i$ (see A5). The energy of any such level can be computed as
\begin{equation}
    E_\Gamma(n)=\frac{1}{|\Gamma|}Tr P_\Gamma \int_{\gamma^\prime}
    z(z-H_n)^{-1}\frac{dz}{2\pi i},
\end{equation}
with $\gamma^\prime$ the contour described in Fig.~2. Simple
manipulations leads to
\begin{equation}\label{Egamma}
E_\Gamma(n)=\epsilon_F^0+\lambda \langle\phi^0_\Gamma, V_n
\phi^0_\Gamma \rangle +\lambda^2 \beta_{\Gamma,\lambda}(n),
\end{equation}
with $\phi^0_\Gamma$ any norm one vector from $P_\Gamma {\cal H}_F$
and
\begin{equation}
\beta_{\Gamma,\lambda}(n)=\frac{1}{|\Gamma|}Tr P_\Gamma
\int\limits_{\gamma^\prime}
z(z-H_0)^{-1}V_n(z-H_n)^{-1}V_n(z-H_0)^{-1}\frac{dz}{2\pi i}.
\end{equation}
We have an upper bound, $\beta_{\Gamma,\lambda}(n)\leq\bar{\beta}$,
with $\bar{\beta}$ independent of $\Gamma$, $\lambda$ or $n$:
\begin{eqnarray}\label{betabar}
\beta_{\Gamma,\lambda}(n)&\leq& 2 \int_{\gamma^\prime}
|z|\|(z-H_0)^{-1}V_n(z-H_n)^{-1}V_n(z-H_0)^{-1}\|\frac{|dz|}{2\pi
}\nonumber \\
&\leq&
\frac{2\gamma_a^2}{\Delta}\int_{\gamma^\prime}|z|\left(1+\frac{|z+a|}
{\Delta}\right)^2\frac{|dz|}{2\pi}.
\end{eqnarray}
Notice also that
\begin{equation}
    \langle\phi^0_\Gamma, V_n \phi^0_\Gamma \rangle=V_\Gamma+\int
    [V_n(x)-V_{n_0}(x)]|\phi^0_\Gamma(x)|^2dx.
\end{equation}
Using the eigenvectors expansion of $(H_0+a)^{-2}$, one can derive
\begin{equation}
    |\phi_\Gamma^0(x)|^2\leq (\epsilon_F^0+a)^2 k_a,
\end{equation}
leading to
\begin{equation}\label{Vgamma}
    |\langle\phi^0_\Gamma, V_n \phi^0_\Gamma \rangle-V_\Gamma|\leq (\epsilon_F^0+a)^2
    k_a \chi \epsilon.
\end{equation}
Returning to Eq.~(\ref{Egamma}), it follows from
Eqs.~(\ref{betabar}) and (\ref{Vgamma}) that
$E_{\Gamma_0}<\epsilon_F^0$ and $E_{\Gamma}>\epsilon_F^0$ for
$\Gamma\neq \Gamma_0$, as long as
\begin{equation}\label{epsilon}
    \epsilon<\frac{|V_{\Gamma_0}|-\lambda
    \bar{\beta}}{(\epsilon_F^0+a)^2 k_a \chi}.
\end{equation}

\begin{figure}
\begin{center}
\includegraphics[width=10.0cm]{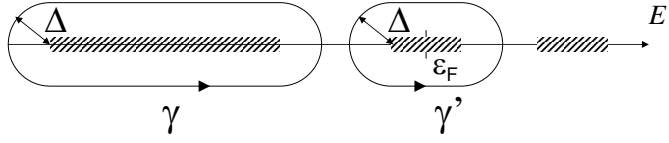}\\
\end{center}
  \caption{The two contours, $\gamma$ and
  $\gamma^\prime$, surround the states below
  $\epsilon_F^0$ and the states split from the Fermi
  level, respectively, such that the distance from any point on the contours to ${\cal
  I}$ is equal to some $\Delta>0$.
  }
\end{figure}

The last thing we need to prove is that, for $\lambda$ small enough,
\begin{equation}\label{invariance}
    \|\textsf{T}_\epsilon[n]-n_0\|_{L^1}\leq \epsilon,
\end{equation}
for any $n$ in $S^{N,\epsilon}_{sym}$ and $\epsilon$ satisfying
Eq.~(\ref{epsilon}). Let us consider
\begin{equation}
    \hat{Z}[n]=\int_{\gamma} (z-H_n)^{-1}\frac{dz}{2\pi i}+
    P_{\Gamma_0}\int_{\gamma^\prime} (z-H_n)^{-1}\frac{dz}{2\pi i},
\end{equation}
defined on the entire $S^N$. Based on the previous results, we
observe that $P_n^<=\hat{Z}[n]$ for $n\in S^{N,\epsilon}_{sym}$,
with $\epsilon$ satisfying Eq.~(\ref{epsilon}). Also, notice that
$n_0(x)=\hat{Z}[\bar{n}](x,x)$. Moreover, if $\|\ \|_1$ denotes the
trace norm,
\begin{equation}\label{fundamental}
    \|\hat{Z}[n]-\hat{Z}[n^\prime]\|_1\leq\frac{\lambda \alpha \chi k_a}{(1-\lambda
    \gamma_a)^2}\|n-n^\prime\|_{L^1},
\end{equation}
with
\begin{equation}
    \alpha=\left (\int_{\gamma}+\int_{\gamma^\prime}\right )\left(
    1+\frac{|z+a|}{\Delta}\right)^2 \frac{|dz|}{2\pi}.
\end{equation}
Indeed, let
\begin{equation}\label{B}
    B\equiv (H_0+a)^{-1}(V_n-V_{n^\prime})(H_0+a)^{-1},
\end{equation}
and $g_{a,z}(t)\equiv (t+a)/(z-t)$. After simple manipulations,
\begin{eqnarray}
    \hat{Z}[n]-\hat{Z}[n^\prime]=\lambda
    \left(\int_{\gamma}\frac{dz}{2\pi
    i}+P_{\Gamma_0}\int_{\gamma^\prime}\frac{dz}{2\pi i}\right)
    g_{a,z}(H_n)\times \nonumber \\
    (1+\lambda(H_0+a)^{-1}V_n)^{-1}B
    (1+\lambda V_{n^\prime}(H_0+a)^{-1})^{-1}g_{a,z}(H_{n^\prime})
\end{eqnarray}
and notice that
\begin{equation}
    \|g_{a,z}(H_n)\|\leq 1+\frac{|z+a|}{\Delta},
\end{equation}
for all $n\in S^N$ and $z\in \gamma$ or $\gamma^\prime$. We can
conclude at this step that
\begin{equation}
    \|\hat{Z}[n]-\hat{Z}[n^\prime]\|_1\leq\frac{\lambda
    \alpha}{(1-\lambda \gamma_a)^2}\|B\|_1.
\end{equation}
If we use the polar decomposition $\Delta V=S|\Delta V|$ of $\Delta
V\equiv V_n-V_{n^\prime}$, and define $A\equiv |\Delta
V|^{1/2}(H_0+a)^{-1}$, then from Eq.~(\ref{B})
\begin{equation}
    \|B\|_1=\|A^\ast S A\|_1\leq \|A^*\|_{HS} \|SA\|_{HS}\leq
    \|A\|_{HS}^2,
\end{equation}
and
\begin{equation}
    \|A\|_{HS}^2=\int |\Delta V(x)|
    (H_0+a)^{-2}(x,x)dx\leq k_a \chi \|n-n^\prime\|_{L^1}.
\end{equation}

With Eq.~(\ref{fundamental}) proven, we can easily end the proof of
i). Indeed, for all $n\in S^{N,\epsilon}_{sym}$,
\begin{eqnarray}
    \|\textsf{T}_\epsilon[n]-n_0\|_{L^1}=
    \|\hat{Z}[n](x,x)-\hat{Z}[\bar{n}](x,x) \|_{L^1} \nonumber \\
    \leq \|\hat{Z}[n]-\hat{Z}[\bar{n}] \|_1 \leq \frac{\lambda \alpha
    \chi k_a}{(1-\lambda \gamma_a)^2}\|n-\bar{n}\|_{L^1}
\end{eqnarray}
and
\begin{equation}
    \|n-\bar{n}\|\leq\|n-n_0\|+\|n_0-\bar{n}\|\leq\epsilon+2N_0.
\end{equation}
Thus, Eq.~(\ref{invariance}) is true if
\begin{equation}\label{lambda}
    \frac{\lambda \alpha \chi k_a (2N_0+\epsilon)}{(1-\lambda
    \gamma_a)^2}\leq\epsilon
\end{equation}
and we remark that Eqs.~(\ref{epsilon}) and (\ref{lambda}) can be
simultaneously satisfied if $\lambda$ is small enough.

ii) Observe that if we take $\lambda$ small so as to satisfy
Eq.~(\ref{lambda}), then
\begin{equation}
    \frac{\lambda\alpha \chi k_a }{(1-\lambda \gamma_a)^2} < 1.
\end{equation}
Then $\textsf{T}_\epsilon$ is a contraction, since $P_n^<$ and
$\hat{Z}[n]$ coincide on $S^{N,\epsilon}_{sym}$ and
\begin{eqnarray}
    \|\textsf{T}_\epsilon[n]-\textsf{T}_\epsilon[n^\prime]\|_{L^1}\leq
    \|P_n^<-P_{n^\prime}^<\|_1 \nonumber \\
    =\|\hat{Z}[n]-\hat{Z}[n^\prime]\|_{1}<\|n-n^\prime\|_{L^1}.
\end{eqnarray}
Since $\textsf{T}_\epsilon$ is a contraction on a closed set, it
must have one and only one fixed point.

iii) It follows immediately if we express $P_n^<$ in terms of the
eigenvectors and notice that, at a fixed point, $n(x)=P_n^<(x,x)$:
\begin{equation}
    H_n \phi_i=\epsilon_i \phi_i, \ \
    n(x)=\sum\limits_{\epsilon_i<\epsilon_F^0}|\phi_i(x)|^2.
\end{equation}
Together with $\int n(x) dx=N$, the above equations are exactly the
KS equations at zero temperature.

\section{Examples}
We consider first one of the simplest examples possible: $2N$
particles on a circle of length $a$. The Kohn-Sham Hamiltonian is
$H_n=-\partial_x^2+V_n$, where $x$ is the coordinate along the
circle. The ground state of $H_0\equiv -\partial_x^2$ is
non-degenerate, while all the excited states are doubly degenerate.
Thus, if we populate the states of $H_0$ with $2N$ particles, we end
up with one particle occupying a double degenerate energy level,
containing the states $a^{-1/2}e^{ik_Fx}$ and $a^{-1/2}e^{-ik_Fx}$
($\epsilon_F^0=k_F^2$).

We now go over the constructions considered in the previous section.
The continuous group $G_c$ are the rotations of the circle and the
subgroup $G$ can be taken as the identity plus the reflection $r:\
x\rightarrow -x$. There are two, one dimensional irreducible
representations of $G$, $\chi_\pm(r)t=\pm t$. The projectors
$P_\Gamma$ ($\Gamma\rightarrow \pm$)  are simply given by
\begin{equation}
    (P_\pm f)(x)=\frac{1}{\sqrt{2}}[f(x)\pm f(-x)].
\end{equation}
They decompose ${\cal H}_F$ in the invariant, 1 dimensional spaces
\begin{equation}
    P_+{\cal H}_F=\{\sqrt{\frac{2}{a}}\cos k_Fx\}, \ \
    P_-{\cal H}_F=\{\sqrt{\frac{2}{a}}\sin k_Fx\},
\end{equation}
each of them providing an irreducible representation for $G$. Either
one of the ($\pm$) representations can chooses as $\Gamma_0$ in A6.
We choose  the (+) representation, in which case,
$n_0(x)=\bar{n}+a^{-1} \cos(2k_Fx)$ and condition A7 reads
\begin{equation}\label{A7}
\int V_{n_0}(x)\cos (2k_Fx)dx<0.
\end{equation}
In LDA, if we approximate $v_{xc}(n_0)\simeq
v_{xc}(\bar{n})+v_{xc}^\prime(\bar{n})(n_0-\bar{n})$, Eq.~(\ref{A7})
reduces to
\begin{equation}
    \hat{v}(2k_F)+v_{xc}^\prime(\bar{n})<0,
\end{equation}
where $\hat{v}$ is the Fourier transform of the two-body
interaction. For QA, Eq.~(\ref{A7}) simply means $\hat{K}(2k_F)<0$.
We will be led to the same conditions on the effective potential if
we choose $\Gamma_0$ to be the anti-symmetric (-) representation.

\begin{figure}
\begin{center}
\includegraphics[width=10.0cm]{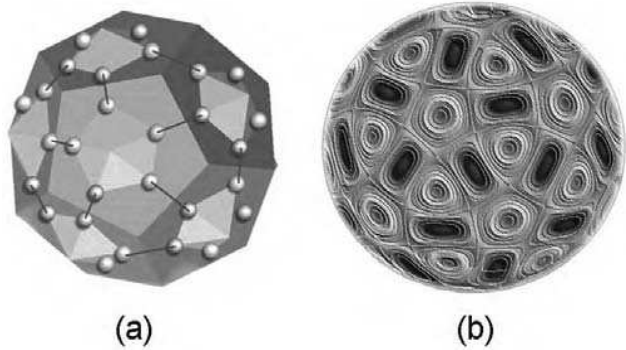}\\
\end{center}
  \caption{a) The molecular structure of the C$_{60}$ molecule. The
  small spheres represent the carbon atoms. The double bonds are
  indicated by segments joining two atoms. b) A contour plot of the density $n_0(x)$
  defined in the text. The dark/lighter regions corresponds to high/low
  densities.
  }
\end{figure}

Similar examples can be given for toruses in higher dimensions. We,
however, consider the case of fermions on the 2D sphere and show
that we can obtain the molecular structure of the C$_{60}$ molecule.
In the C$_{60}$ molecule, the carbon atoms sit at the points of
intersection between an icosahedron and dodecahedron as shown in
Fig.~3a. There are single and double bonds between the carbon atoms.
Since the double bond is much stronger than the single bond, we
consider C=C as being the building blocks of the C$_{60}$ molecule.
In total, there are 30 double bonds and some of them are shown in
Fig.~3a. We consider then 30 point particles (of course, we tried 60
particles with no success) on a sphere of radius $R$, described by
the Kohn-Sham Hamiltonian $H_n=-R^{-2}\vec{L}^2+V_n$, where
$H_0=-R^{-2}\vec{L}^2$ is the kinetic energy of a particle on the
sphere. The energy levels of $H_0$ are simply $R^{-2}l(l+1)$, $l=0,\
1,\ldots$, and if we populate them in order, we end up with 5
particles on the $l=5$ level. The continuous group is $O(3)$ and we
can take the proper icosahedral group as the finite subgroup $G$.
Indeed, under its action, the Fermi level decomposes as
\begin{equation}
    {\cal H}_F=H_u\oplus T_{1u}\oplus T_{2u},
\end{equation}
i.e. in a 5 dimensional space plus two 3 dimensional spaces (of
different symmetries). Thus, the proper icosahedral group satisfies
A5-A6 with $\Gamma_0=H_u$. Then, if the effective potential
satisfies A7 (which we expect to happen for certain radiuses $R$),
the Kohn-Sham equations for the 30 particles on the sphere have a
stable solution $n(x)=n_0(x)+o(\lambda)$, where $n_0(x)$ is obtained
by populating all $l<5$ levels plus the 5 states with $l=5$ and
$H_u$ symmetry. This density is shown in Fig.~3b and the resemblance
with the C$_{60}$ molecule is evident.

We end this section with a discussion of the numerical results of
Ref.~\cite{Reimann} on 2D electrons and a completely deformable
jellium, whose electrostatic potential cancels exactly the
electrons' Hartree potential. This system is described by the
Kohn-Sham Hamiltonian:
\begin{equation}\label{KS2D}
    H_n=-\Delta+\bar{v}_{xc}(n),
\end{equation}
where $\bar{v}_{xc}(n)$ is an effective 2D exchange-correlation
potential (treated in the local density approximation). From the
beginning, we should mention that this system does not fall in our
category. We discussed here systems that are confined in finite
volumes and the small coupling constant regime. In Eq.~(\ref{KS2D}),
the electrons are free to move in the entire 2 dimensional space.
The only confining potential is their own exchange-correlation
potential. Since this potential needs to bind these electrons, we
cannot talk about the small coupling limit. However, we will show
that our analysis still applies. We mention that a group analysis of
the symmetry breaking in small parabolic quantum dots was already
carried in Ref.~\cite{Uzi}.

The self-consistent numerical solutions of Eq.~(\ref{KS2D}) showed
the following: there is a first class of stable crystals, which have
pure shapes, like triangles, squares and circles, and there is a
second class with apparently no regularity in the shape. We discuss
the crystals with pure shape, where the symmetry group is easy to
recognize, and the case when the electrons are paired and the spin
degree of freedom is not important. Clearly, for the triangle-shaped
crystals, the symmetry has been broken to $G=C_3v$, the group that
sends an equilateral triangle into itself. For the square ones,
$G=D_{4v}$, the group that sends a square into itself. In
Ref.~\cite{Reimann}, it was found numerically that crystals with 8
and 10 particles prefer the square geometry. Let us show how one can
predict this. We first guess a good density $n^{(0)}$, which is a
uniform density inside a square.  $n^{(0)}$ will create a constant,
(strongly) negative potential inside the square and the electrons
will behave, more or less, like they are confined within hard walls.
The eigenstates for this system are
$\psi_{nm}(x,y)=\frac{2}{A}\sin(n\pi x/L)\sin(m\pi y/L)$ ($L$ the
size of the square and $A=L^2$). The lowest energy state is
$\psi_{11}$. The first excited state is double-degenerate and
corresponds to $\psi_{12}$ and $\psi_{21}$: $D_{4v}$ acts
irreducibly on this space. The third excited state is non-degenerate
and correspond to $\psi_{22}$. The fourth excited state is again
double degenerate and corresponds to $\psi_{13}$ and $\psi_{31}$:
under the representations of $D_{4v}$, this level decomposes into 2
invariant, 1 dimensional spaces (symmetric and anti-symmetric
combinations of $\psi_{13}$ and $\psi_{31}$). Thus, with 8
particles, we can populate the first 3 levels and have the closed
shell condition. Thus, the self-consistency can be achieved without
further reduction of the symmetry. If we write
$H_0=-\Delta+\bar{v}_{xc}(n^{(0)})$, we reduced the Kohn-Sham
Hamiltonian to $H_n=H_0+V_n$, where $H_0$ describes, more or less,
free electrons confined in the square and
$V_n=\bar{v}_{xc}(n)-\bar{v}_{xc}(n^{(0)})$ can be considered small.
The density $n_0$ of our theorem is given by populating the first 3
energy levels of $H_0$ and proving the existence of a
self-consistent solution near $n_0$ may be accomplished now with the
methods developed in this paper. The self-consistent field will
split the fourth energy level, so we can accommodate another pair of
electrons without breaking the symmetry, i.e. the crystal with 10
electrons should also have square symmetry. Examining either Fig.~2
or the inset of Fig.~3 (where one can see the splitting of the
second energy level) in Ref.~\cite{Reimann}, it follows that, in
reality, there is a slight deviation from the $D_{4v}$ symmetry.

\section{Discussion}

We can summarize our procedure as follows (hopefully this can be
seen in the above examples): in the first part, one looks for $n_0$,
which must be a good approximation of the self-consistent density.
The main goal is to fulfill the closed shell condition and group
theoretical methods can be used to accomplish that. After $n_0$ is
found, one can construct the map $\textsf{T}$ in a neighborhood of
$n_0$. In the second part, one investigates if $\textsf{T}$ is
indeed a contraction near $n_0$. For small $\lambda$, the conditions
in which this is true are already given here. How large this
$\lambda$ can be depends on $V_n$ and its derivative near $n_0$.

One important remark is that some of these non-uniform solutions can
be found only if we start the iteration close enough to them. The
reason is that the basin of attraction of the map $\textsf{T}$ may
be small (notice that the usual iterative process of solving KS
equations consists exactly in constructing  the sequence $n_0$,
$\textsf{T}[n_0]$, $\ldots$, $\textsf{T}^m[n_0]$, etc.). For this
reason, there may be additional isomers to the ones found in
Ref.~\cite{Reimann}.

As opposed to the Jahn-Teller effect \cite{Jahn} or Pierls
instability \cite{Peierls}, which involves discrete symmetries and
the electronic degeneracies are lifted by a displacement of the
atoms, the symmetry breaking discussed here involves a continuous
group and is due solely to the electron-electron interaction, which
lifts the electronic degeneracies without any change in the external
potential. The Jahn-Teller and Pierls instabilities may be triggered
(but not necessarily) by the instability we discussed here.

Our analysis does not rule out the existence of more than one
self-consistent solution of the KS equations. The lowest energy
configuration is, of course, associated with the ground state, and
the higher energy configurations should be associated with excited
states. In Ref.~\cite{Levy}, the authors find that indeed, since the
energy functional $E_v[n]$ is not convex, $E_v[n]$ may have
additional extrema, which are excited-state densities.

Degeneracy and symmetry breaking in DFT are well studied concepts in
Density Functional Theory \cite{Savin1, Savin2}, and there are many
numerical studies on the subject. In fact, the modern theory of
freezing \cite{Baus}, which can be traced back to the pioneering
work of Ramakrishnan and Yussouff \cite{Yussouff}, is based on the
assumption that the liquid-solid transitions occurs because of a
bifurcation of the same type as we discussed here. More exactly, the
density of the solid $n_s$ is computed self-consistently as the
linear response of the uniform liquid density $n_l$ to the
introduction of a density change $\Delta n=n_s-n_l$. In the current
state, the procedure cannot predict the lattice symmetry, but rather
assumes that the linear response equation has a non-uniform
solution, with a prescribed crystalline symmetry (usually known from
experiment). Applied to quantum liquids, this procedure is
equivalent to solving the Kohn-Sham equations in the quadratic
approximation. One can find in Ref. \cite{Senatore} an impressive
numerical demonstration of the Wigner crystallization of the
electron liquid. For finite systems, we already mentioned
Refs.~\cite{Reimann} and \cite{Uzi}. We add Refs.~\cite{Filinov} and
\cite{Yannouleas}.

We also want to mention that Bach et al \cite{Lieb} have shown that,
for any repulsive interaction, the energy levels are always fully
occupied in the unrestricted Hartree-Fock approximation. This result
automatically implies that there must be a symmetry breaking
whenever the last occupied level of $H_0$ is only partially
populated. In contradistinction, Ref.~\cite{Prodan3} showed that,
within the Hartree approximation (1D), there is always a symmetry
breaking for short, attractive interactions. This definitely shows
that we have to go beyond the two approximations.

At the end, we want to mention that we have partial but interesting
results in the thermodynamic limit and we are currently considering
the finite temperature regime. We are also looking into the spin
case, when the Kohn-Sham equations become self-consistent equations
for the density $n(x)$ and magnetization {\it vector} $\vec{m}(x)$.

\end{document}